# Development of a Hard X-ray focal plane Compton Polarimeter: A compact polarimetric configuration with Scintillators and Si photomultipliers

T. Chattopadhyay · S. V. Vadawale · S. K. Goyal · Mithun N. P. S. · A. R. Patel · R. Shukla · T. Ladiya · M. Shanmugam · V. R. Patel · G. P. Ubale

**Abstract** X-ray polarization measurement of cosmic sources provides two unique parameters namely degree and angle of polarization which can probe the emission mechanism and geometry at close vicinity of the compact objects. Specifically, the hard X-ray polarimetry is more rewarding because the sources are expected to be intrinsically highly polarized at higher energies. With the successful implementation of Hard X-ray optics in NuSTAR, it is now feasible to conceive Compton polarimeters as focal plane detectors. Such a configuration is likely to provide sensitive polarization measurements in hard X-rays with a broad energy band. We are developing a focal plane hard X-ray Compton polarimeter consisting of a plastic scintillator as active scatterer surrounded by a cylindrical array of CsI(Tl) scintillators. The scatterer is 5 mm diameter and 100 mm long plastic scintillator (BC404) viewed by normal PMT. The photons scattered by the plastic scatterer are collected by a cylindrical array of 16 CsI(Tl) scintillators (5 mm×5 mm×150 mm) which are read by Si Photomultiplier (SiPM). Use of the new generation SiPMs ensures the compactness of the instrument which is essential for the design of focal plane detectors. The expected sensitivity of such polarimetric configuration and complete characterization of the plastic scatterer, specially at lower energies have been discussed in [11,13]. In this paper, we characterize the CsI(Tl) absorbers coupled to SiPM. We also present the experimental results from the fully assembled configuration of the Compton polarimeter.

T. Chattopadhyay
Astronomy and Astrophysics Division, Physical Research Laboratory, Ahmedabad, India

Indian Institute of Technology, Gandhinagar, India
Tel.: +91-079-26314617
E-mail: tanmoy@prl.res.in

S. V. Vadawale · S. K. Goyal · Mithun P. S. · A. R. Patel · R. Shukla · T. Ladiya · M. Shanmugam · V. R. Patel · G. P. Ubale
Astronomy and Astrophysics Division, Physical Research Laboratory, Ahmedabad, India

**1 Introduction**

X-ray polarimetry promises to open a new window in Astrophysics with its vast applications all the way from distant AGNs and blazars, galactic black hole and neutron star X-ray binaries to the solar atmosphere. However, the field of X-ray polarimetry has been mostly unexplored since the birth of X-ray astronomy in early 1960s. Initially, in the 1970s, there were many attempts and initiatives [1,54,39,23,22,52,53,44,25] in the measurements of X-ray polarization for celestial objects. However, in the next 25 years, there was no major advancement in the field of X-ray polarimetry except for few mission proposals based on Bragg and Rayleigh polarimeters [28,46,48,18,34]. The main reason behind the lack of progress in X-ray polarimetry is the high photon throughput which results in poorer sensitivity of the X-ray polarimeters.

With the vast improvement in the detection technology in the last decade or so, particularly with the invention of photoelectric polarimeters [16,3,5,6], X-ray polarimetry has witnessed a significant growth in interest of the astronomical community. Few polarimetric missions were proposed based on the photoelectric polarimeters [15,47,26,4,45]. Among them, Gravity and Extreme Magnetism Small Explorer (GEMS) [26], carrying a Time Projection Chamber (TPC) based photoelectric polarimeter [6] was the only dedicated X-ray polarimetry mission selected for launch in 2014. However, the mission was cancelled due to programmatic issues. Though photoelectric polarimeters are expected to provide sensitive polarization measurements, these instruments are effective primarily in soft X-rays where radiation from the source is expected to be less polarized because of the dominance of thermal radiation. Consequently, many groups across the globe are now involved in developing Compton polarimeters effective in hard X-ray regime where the expected polarization is above the typical sensitivity level of the instruments [56,31,29,8,40]. For the same reason, there have been several attempts during the last decade in retrieving polarization information from instruments, not specifically designed for polarimetry but could be sensitive to it [35,14,33,27,19,17,38,21,20,36,37,32,12,50]. The major issue with such attempts is the non-optimized polarimetric configuration of these instruments which results in quite poor sensitivity. Hence such bonus polarization measurements are limited only for few very bright and highly polarized sources like Crab, CygX1, or for Gamma Ray Bursts.

With recent development of Hard X-ray optics e.g. NUSTAR, Astro-H, now hard X-ray polarimetry may see manyfold improvement in terms of sensitivity of the polarimeters. Compton polarimeters at the focal plane of hard X-ray telescopes are expected to provide sensitive polarization measurements because of two factors − 1. compact focal plane detectors can be designed with an optimized configuration for polarimetry, and 2. concentration of flux in hard

X-rays and narrow FOV of the telescopes reduces the background which significantly improves the sensitivity of the focal plane polarimeters. Motivated by this, we are developing a focal plane Compton X-ray polarimeter (CXPOL) for hard X-ray optics [11,13]. The planned polarimetric configuration consists of a long thin plastic scintillator surrounded by a cylindrical array of CsI(Tl) scintillators viewed by SiPMs. The development of the polarimeter is essentially an extension of our study of the Rayleigh polarimeter [51] for a proposed ISRO mission called POLIX [42]. The other objective of the study is also to show our readiness level prior to proposing for a future hard X-ray polarimetry mission. In our earlier work [11], we estimated polarimetric sensitivity of the instrument using detailed Geant4 simulations followed by complete characterization of the central plastic scatterer [13]. In this paper, we start with a brief description of our earlier work in section 2, followed by a detailed experimental study on characterization of CsI(Tl) absorbers in section 3. In section 4, we describe the polarization experiment with CXPOL using polarized sources of different energies and finally, we conclude with future plans regarding the optimization of the instrument.

## 2 CXPOL configuration

The proposed focal plane Compton polarimeter configuration consists of a 5 mm diameter and 10 cm long plastic scintillator as central scatterer surrounded by a cylindrical array of 16 CsI(Tl) scintillators (see Fig. 1). The CsI(Tl) ab-sorbers are 15 cm long and 5 mm × 5 mm in cross-section. We carried out detailed Geant4 simulation for this polarimeter configuration in 20−80 kev, 80 kev being the upper energy detection limit of hard X-ray optics (assumed to be the NUSTAR type optics for this study). Minimum Detectable Polarization (MDP) of the instrument was estimated taking into account the effective area of NUSTAR optics in 3−80 kev and proper background estimates. Background chance events within a coincidence time window of 10 $\mu$s was found to be insignificant which is expected for a Compton polarimeter thanks to the active coincidence between the scatterer and absorbers and narrow FOV of hard X-ray telescopes. The MDP for a 100 mCrab source was found to be $\sim$ 1 % for 1 Ms exposure, whereas for 100 ks exposure, the expected sensitivity of the instrument is $\sim$ 3 %. Details of simulation and polarimetric sensitivity estimations are reported in [11]. Here it is to be noted that reflection off the hard X-ray mirror may introduce small artificial modulation in the azimuthal angle distribution due to the dependence of mirror reflectance on the polarization plane. Recent studies [30] suggest that change in polarization because of the hard X-ray mirrors is around $\sim$ 1 % limiting the sensitivity of the polarimeters at the focal plane of such hard X-ray telescopes to $\sim$ 1 %. There are few other polarimetric mission proposals based on hard X-ray focal plane Compton polarimeters like X-Calibur [2], PolariS [24] and TSUBAME [55]. Among these, our polarimetric configuration closely resembles with the scattering geometry used in X-Calibur [2], which is under active consideration for NASA's next small satellite mission PolSTAR. The basic difference between these two configurations is that X-Calibur employs CZT detector array as absorbers which also enables the feasibility of spectro-polarimetry in 20−80 kev. However, use of scintillators with SiPM readout makes our scattering geometry more optimized for polarimetry measurements, comparatively less complex electronically and much more compact.

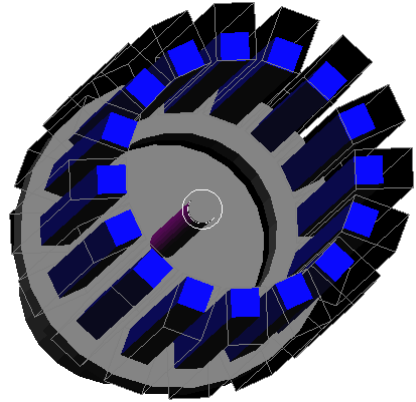

**Fig. 1** Planned Compton polarimeter configuration. The photons scattered by the central plastic scatterer (10 cm long with 5 mm diameter) are collected by a cylindrical array of 16 CsI(Tl) scintillators (each 15 cm long and 5 mm × 5 mm cross-section). In [11], we reported simulation for the Compton polarimeter geometry assuming 32 absorbers; however in the current experimental configuration, 16 CsI(Tl) scintillators have been used.

Polarimetric sensitivity of this kind of configuration critically depends on the lower energy threshold of the active plastic; lower the threshold of plastic, better is the sensitivity. Therefore, it is important to characterize the behavior of the plastic particularly at lower energies to estimate the realistic polarimetric sensitivity. However, due to various statistical processes involved in the detection of events in this kind of detectors, we do not expect a sharp threshold in the plastic scintillator. Since it is difficult to estimate the threshold for an organic scintillator with usual spectroscopic methods because of their poor energy resolution, we carried out an experiment in which the lower energy threshold of the plastic scatterer was measured in Compton mode. Details of the experiment set up and results can be found in [13]. Overall it was observed that the detection efficiency (i.e. probability that a trigger will be generated at the output of plastic for a given energy deposition) of our plastic scatterer configuration was $\sim$ 5 % for the deposited energy of $\sim$ 0.5 keV. The efficiency increased with the increase in deposited energy in plastic and reached 100 % at energies $\sim$ 7 kev. The lower energy threshold for polarization with this scatterer was found to be $\sim$ 14 kev. Though these results are applicable for the particular scatterer configuration (plastic scintillator geometry, PMT, coupling etc.) used in our experiment, in general it is expected that an active plastic scatterer will result in similar (or even better for a shorter scatterer) efficiency for lower energy deposition. Inclusion of the measured plastic detection efficiency in the MDP calculation results in much more realistic polarimetric sensitivity of the instrument. MDP for a 100 mCrab source is found to be $\sim$ 2 % for 1 Ms exposure, whereas sensitivity degrades to $\sim$ 4 % for 100 ks exposure.

## 3 Characterization of CsI(Tl) scintillators

In our final configuration of the Compton polarimeter, the scattered photons from plastic scatterer are collected by 16 CsI(Tl) scintillators. Each of the CsI(Tl) crystal is 15 cm long and 5 mm $\times$ 5 mm cross-section (see Fig. 2), procured from Saint-Gobain. The CsI(Tl) crystals are known to have high

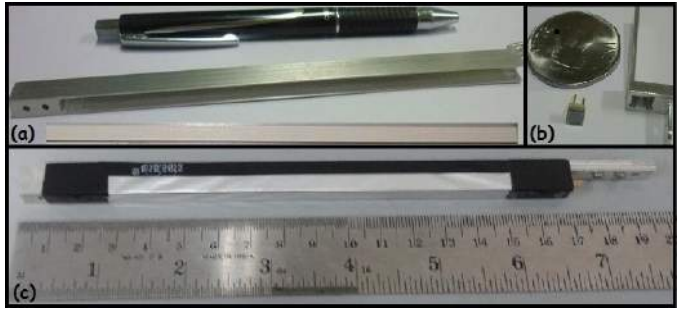

**Fig. 2** (a) A CsI(Tl) scintillator (5 mm $\times$ 5 mm $\times$ 150 mm) and its aluminium holder. (b) A Si photomultiplier (3 mm $\times$ 3 mm) with its aluminium case. (c) The complete CsI(Tl)-SiPM package. The CsI(Tl) is coupled to the SiPM using optical glue and wrapped by a thin aluminium foil for light tightness. 16 such CsI(Tl)-SiPM systems have been used in the final Compton polarimeter configuration.

light yield, however, for our initial experiment these crystals are selected primarily because of their less hygroscopic nature compared to other well known inorganic scintillators such as NaI(Tl) or LaBr$_3$(Ce), which makes them easier to handle in the laboratory. One important drawback of CsI(Tl) scintillator is the long scintillation decay time, which is particularly severe in the context of readout by SiPM. Thus our results with CsI(Tl) readout by SiPM are expected to be worst case scenario and could be significantly improved with faster scintillators.

Each of the CsI(Tl) crystals are kept inside an aluminium case which encloses the crystal from all sides except the side facing the scatterer and the lower end for readout. The aluminium case is 5 mm thick on the back and 1 mm thick on the both sides. Each of the scintillators is read by a single Si photomultiplier (SiPM) at one end of the CsI(Tl). SiPM is a multi-pixel semiconductor photodiode system with pixels on a common Si substrate. SiPM work in Geiger mode (bias voltage > breakdown voltage), which provides high gain ($\sim 10^6$) making it sensitive even for single photon detection. The details of working principle of SiPM and other properties can be found in [10,9,41].

We used SiPM procured from KETEK, Germany (SiPM PM3350), with active area of 3 mm × 3 mm (see Fig. 2). The device is small, light, and robust with low operation voltage ($\sim 31$ V) and therefore easier to handle and provides the compactness necessary for focal plane detectors. The wide spectral range from 300 nm to 800 nm (peak wavelength 420 nm) nicely matches with the CsI(Tl) emission spectra. There are 3600 micro-pixels (each of 50 $\mu$m × 50 $\mu$m) in a single device. An incident photon on any micro-pixel triggers an avalanche. In low light conditions, the number of fired micro-pixels is directly proportional to the number of photons incident on entire active surface. Hence in such conditions the SiPM can be used to measure the intensity of incident light by adding the signals from all the fired micro-pixels, though individual micro-pixels are operating in Geiger mode. One major problem with SiPM is the constant leakage current resulting from the random firing of micro-pixles due to thermal and field excitations inside Si, which makes it difficult for lower energy applications. At energies beyond 100 kev, SiPM device has been proved to be a much better readout option for scintillators than conventional vacuum PMTs [7,43]. Here we plan to use SiPM to read out CsI(Tl) crystals at energies below 100 kev, which essentially depends on many factors like background level in the SiPM (typically $\leq 500$ kHz/mm$^2$), good coupling between the crystal and SiPM, lower electronic noise, and on the scintillator properties (good light collection efficiency and small decay constant). An aluminum holder is used to keep the SiPM and couple it at the end of the CsI(Tl). For better coupling, optical glue with suitable refractive index is used between CsI(Tl) and SiPM. The whole system is then wrapped by thin aluminium foil from all sides to make it light tight (see Fig. 2).

The front end electronics for a single CsI(Tl)−SiPM system readout is shown in Fig. 3. The output of the SiPM is read by a charge sensitive pre-amplifier (RC type CSPA) utilizing ac coupling between CSPA and SiPM. The signal, in this way, is always detected at the top of the constant background leakage current. The output of the CSPA is given to a pulse shaping amplifier (CR−RC−RC), with 3 $\mu$s peaking time. The shaped pulse output is accepted by a multi channel Analyser (MCA) to record spectra from CsI(Tl). 16 similar electronic chains are developed for read out of 16 CsI(Tl) bars. All 16 scintillators as well as electronic chains are not expected to have absolutely identical characteristics because of factors like unequal coupling between scintillators and SiPM and variation in gain across the electronics chains. Therefore, we characterized each of the 16 CsI(Tl)-SiPM detectors with multiple radio active sources ($^{241}$Am, $^{109}$Cd, $^{57}$Co) in the energy range of 20−130 keV, which is also the polarimetric energy range of the instrument. The experiment setup is shown in Fig. 3.

Fig. 4 shows the spectra for $\sim 26$ kev and 59.54 kev lines ($^{241}$Am) in black, 22 kev and 88 kev lines ($^{109}$Cd) in red and 122 kev line ($^{57}$Co) in blue obtained from one of the CsI(Tl) scintillators. The presence of 22 kev line clearly shows that lower energy threshold of CsI(Tl)-SiPM is $\sim 20$ kev which is essential as we plan a lower energy cut off of $\sim 20$ kev for the Compton polarimeter. However, it is to be noted that spectra shown in Fig. 4 is obtained when the sources are kept close to SiPM. For interactions far away from SiPM, we expect less light to reach SiPM device and because of reflections inside the scintillator, light signal is expected to be diffused as it reaches the SiPM plane. Therefore, we investigate the response of CsI(Tl) as a function of distance from SiPM by changing the source position with a step of 1 cm. The sources are kept very close to the CsI(Tl) to make sure that interaction takes place in a very small region in CsI(Tl). For each source position, we acquire the spectra and fit the individual lines with Gaussian profile to estimate the peak channel of the lines. The fitted peak channels are plotted as a function of distance as shown in Fig. 4. We find that the peak channel (ADC) of detection falls off exponentially with distance from SiPM. The dashed red line at $200^{th}$ ADC channel indicates the background level in the spectra. We see that for 59.5 kev photons from $^{241}$Am, the sensitivity degrades significantly beyond 4 cm. Similarly for 88 kev and 122 kev photons, the effective length of CsI(Tl) is $\sim$ 5 cm and 6 cm respectively. The steep fall in detection probability is mainly because of diffusion of light signal each time the photons undergo reflections inside CsI(Tl). Though, better response is expected with better optical coupling between CsI(Tl) and SiPM, these results indicate that CsI(Tl) scintillators as long as 15 cm are not suitable as absorbers in Compton polarimeter particularly when viewed by a single SiPM at one end. However, we expect better performance in case of faster scintillators like LaBr$_3$(Ce), CeBr$_3$ etc. with light output similar to that of CsI(Tl) as even for distant interactions, number of photons reaching SiPM at an instant is higher compared to that of slower scintillators like CsI(Tl). As discussed earlier, such a steep fall in detection probability with distance is expected for slow scintillators like CsI(Tl) particularly with SiPM readout. In spite of that, we selected CsI(Tl) scintillators for this proof of concept experiment mainly due to the fact that these are less hygroscopic compared to the other inorganic scintillators and therefore comparatively easier to handle in the laboratory. In the later versions of the polarimeter, we plan to use faster scintillators with SiPM readout as absorbers. We also verify for the linearity of the detectors in this energy range for different source positions to estimate gain and offsets. Scintillator to scintillator variation in gain for a fixed position is found out to be insignificant, which is important for polarimetry applications.

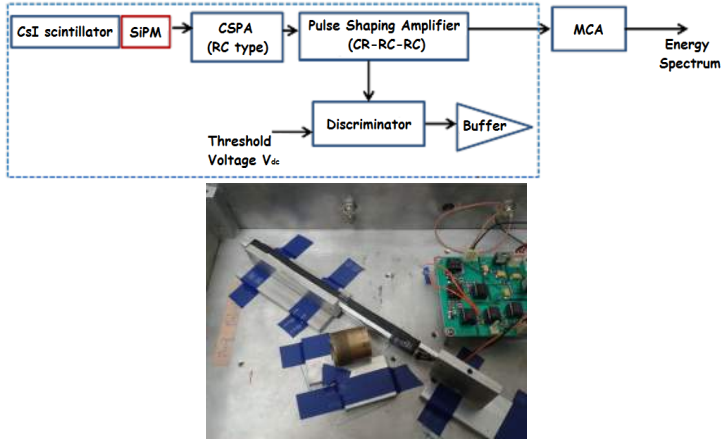

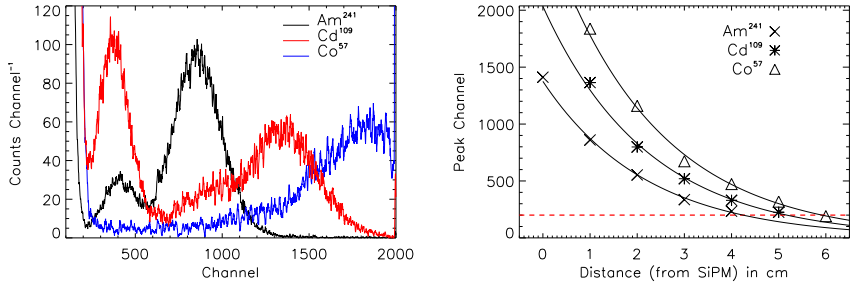

**Fig. 3** Top: Schematic of the SiPM electronic readout consisting of CSPA, pulse shaper and MCA. Bottom: Experiment set up to characterize CsI(Tl)-SiPM system.

**Fig. 4** Left: Spectrum obtained from a CsI(Tl)-SiPM system for 26 and 59.5 kev photons from $^{241}$Am (black), 22 and 88 kev photons from $^{109}$Cd (red), and 122 kev photons from $^{57}$Co (blue). Right: Detection probability of CsI(Tl) as a function of distance from the SiPM for 59.5 kev (cross), 88 kev (asterisk) and 122 kev photons (triangle). The solid lines are the exponential fit to the experimental data. The red dashed line denotes the typical background level in the spectrum in ADC channel unit.

## 4 Polarization experiment with CXPOL

After characterizing each of the 16 CsI(Tl) scintillators, all of them are integrated on the Compton polarimeter housing in the form of cylindrical array to test the performance of the polarimeter with polarized radiation beam. The supporting structure to hold the absorbers surrounding the central plastic is made of aluminium. It should be noted that the simulations results, reported in [11], were carried out with total 32 absorbers having slightly larger diameter (5.3 cm) of the cylindrical array, whereas the current polarimeter configuration has diameter of 4 cm. The modulation pattern is not expected to change significantly with 16 absorbers because the azimuthal bin sizes are all equal. Position of the plastic can be altered within the polarimeter structure. In the original polarimeter configuration, the plastic scatterer was planned to be kept at the base of the polarimeter, making full length of the scatterer available for the coincident detection. However, since the CsI(Tl) scintillators were found to be sensitive only at the lower $\sim 5-6$ cm at energies $\sim 50$ kev, the plastic scatterer was pushed out by 5 cm to ensure that scattered photons from the top portion of the scatterer interact within the active length of CsI(Tl).

The coincidence between the plastic and the CsI(Tl) absorbers is established utilizing the gated mode operation using FPGA. The schematic of the coincidence unit is shown in Fig. 5. The output of the first stage amplifier in the SiPM readout chain is given to a discriminator, which compares the output with a fixed threshold voltage and gives signal for the presence of a photon. The fixed voltage is optimized to reject a major fraction of noisy signals. All 16 discriminator outputs and the discriminator output of the central plastic scintillator are given to the FPGA based counting system. All the 17 signals are counted on the rising edge of the clock of FPGA. When there is a signal from the plastic scintillator, FPGA records the presence of all other 16 Compton scattered signals for $\sim 6$ $\mu$s coincidence time window. Based upon the coincident detection (within 6 $\mu$s window) of signal from the plastic scatterer and XORed signal of the 16 CsI(Tl) signals (to avoid multiple scattering due to fluorescence photons from CsI(Tl)), the detected signals are sent to a LabVIEW data acquisition software which takes the packet data from the FPGA and store in an output file for further analysis.

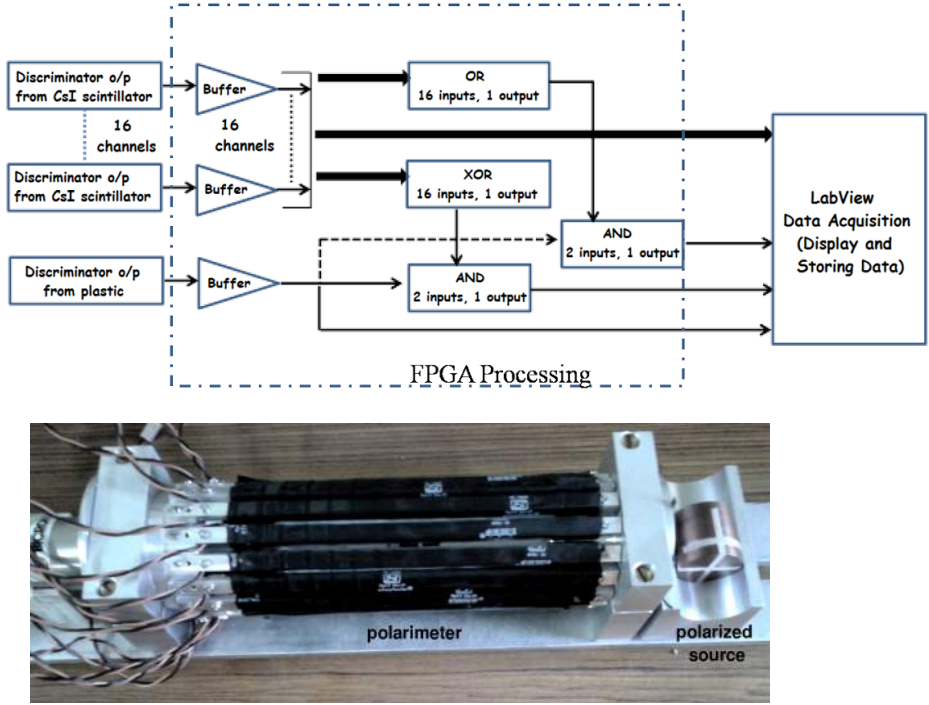

**Fig. 5** Top: Schematic of the coincidence unit between the plastic scatterer and the CsI(Tl) absorbers. Bottom: Polarization experiment set up with the fully integrated configuration of the Compton polarimeter. The polarized source of radiation (shown in the figure) employs 90° Compton scattering of the unpolarized photons from radioactive sources (see text for details).

Fig. 5 shows the polarimetric configuration and experiment setup to investigate the response of CXPOL to unpolarized and polarized beam of radiation. The source is kept outside the polarimeter case. In order to make sure that the photons do not diverge and impinge directly on the CsI(Tl) scintillators, we used a long lead-aluminium collimator at that end. The azimuthal distribution for an unpolarized beam is shown in Fig. 6. The plastic scatterer is exposed to 59.5 kev photons from $^{241}$Am. We see that there is no significant modulation in the azimuthal angle distribution. The distribution is fitted with a $\cos^2 \phi$ function (shown by black line) —

$$C(\phi) = A \cos^2(\phi - \phi_0) + B \quad (1)$$

where A, B, and $\phi_0$ (angle of polarization) are the fitting parameters. Amplitude of modulation in the azimuthal angle distribution is given by modulation factor

$$\mu = \frac{A}{A + 2B} \quad (2)$$

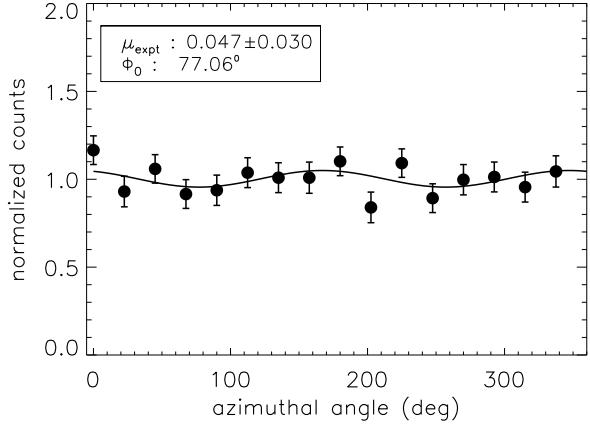

**Fig. 6** Azimuthal angle distribution for unpolarized 59.5 kev photons. The distribution is fitted by a $\cos^2 \phi$ function (see Eq. 1) shown by black line.

Modulation factor is directly proportional to the polarization fraction of the beam. The small nonzero modulation in this case is because of slight difference in gains between the scintillators.

We tested the response of CXPOL to a partially polarized beam of 54 kev, obtained by 90° scattering of 59.5 kev photons from $^{241}$Am. An aluminium rod was used as scattering element. Both the scatterer and the source were kept inside a thick lead cylinder with a 2 mm diameter hole perpendicular to the incident beam direction. Spectrum of the polarized beam as taken by a separate CdTe (Amptek X123CdTe) detector is shown at the top of the left column in Fig. 7. The broad peak centered around 54 kev is the scattered polarized beam. The narrow beam at $\sim$ 59.5 kev is the Rayleigh scattered photons of the original 59.5 kev photons from $^{241}$Am. We estimated the polarization fraction of the beam numerically. The overall polarization will depend on the energy of the incident photons and the geometry of scattering of the source. The fraction of polarization turns out to be ∼ 75 %. When this partially polarized beam is incident on the CXPOL, we see a clear enhanced modulation for the polarized beam compared to the unpolarized case at 0° and 90° polarization angles (see left column of Fig. 7). Different polarization planes are achieved by rotating the lead cylinder with respect to the plastic axis. First we take background data for long exposure which is then subtracted from source data.

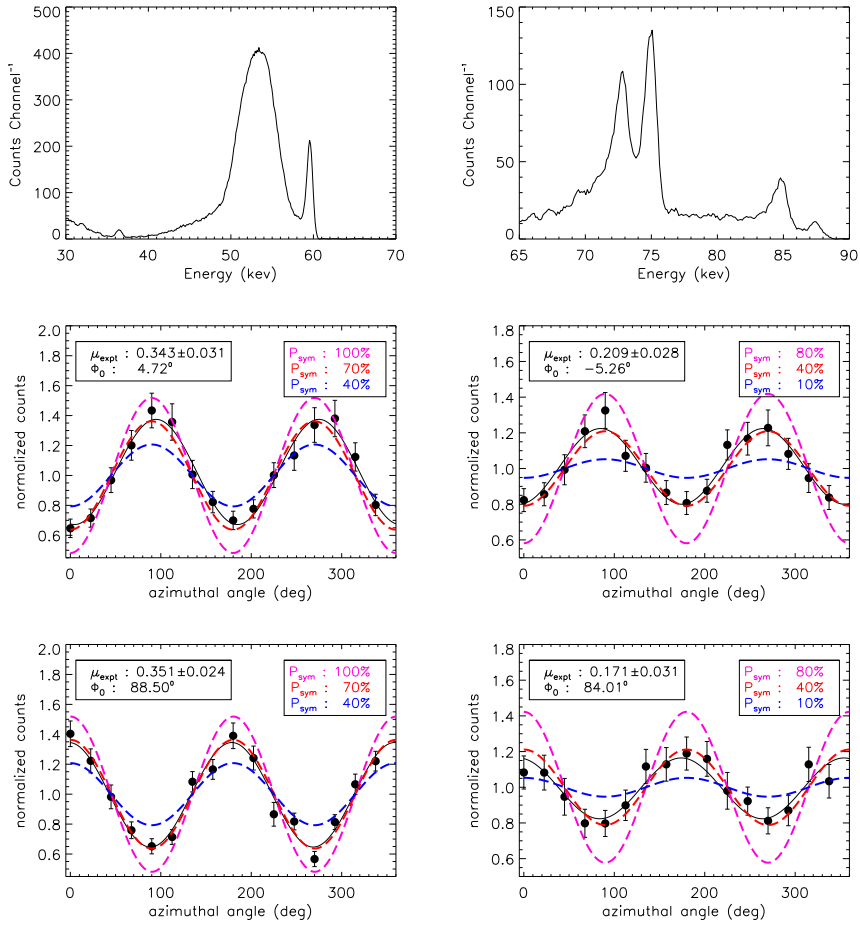

**Fig. 7** Left column: Top - Spectrum (taken from CdTe detector) of polarized 54 kev beam obtained by 90° scattering of 59.5 kev photons from $^{241}$Am. Middle and Bottom - Azimuthal angle distribution for the partially polarized 54 kev beam for polarization angles of 0° and 90° respectively. The solid black line is the fit to the experimental data, whereas the dashed lines are obtained from simulation for the setup for polarization fractions of 100 % (pink), 70 % (red) and 40 % (blue). Right Column: similar to the left column, obtained from $^{109}$Cd. In this case, the amplitude of modulation is less due to the presence of unpolarized lead fluorescence photons (72 and 84 kev) as shown in the spectrum (see text for details).

The modulation patterns are fitted with $\cos^2\phi$ function (see Eq. 1) shown by solid black line. Modulation factors for both the polarization angles are found to be $\sim 0.35$. To compare the fitted modulation factors with simulation results, we performed Geant4 simulation for the current configuration of the polarimeter. Modulation factor for 100 % polarized 54 kev beam ($\mu_{100}$) is found to be $\sim 0.50$. The conventional way to obtain the degree of polarization of any partially polarized beam is to take ratio of the experimentally obtained modulation factor $\mu_{expt}$ to the simulated $\mu_{100}$ i.e.

$$P = \frac{\mu_{expt}}{\mu_{100}} \quad (3)$$

This results in a polarization fraction of $\sim 70 \pm 8$ %, which is in good agreement with the numerically estimated value for the 54 kev polarized beam. This is also demonstrated in Fig. 7, where the modulation curves obtained from simulation are shown in dashed lines, where pink, red, and blue denote 100 %, 70 %, and 40 % polarization respectively. The experimental data agree well with the 70 % polarized signal, as expected.

We repeated the same experiment with partially polarized beam from $^{109}$Cd source. The plot at the top of the right column of Fig. 7 shows the spectrum of this beam centered around 75 kev (90° scattering of 88 kev photons from $^{109}$Cd). These 88 kev photons induce fluorescence emission (72 kev and 84 kev) from the surrounding lead enclosure as seen in the spectrum. These unpolarized photons are expected to decrease the polarization fraction of the beam. Eventual polarization degree is expected to be $\sim 38$ % in this case, estimated taking into account the area under the 72, 84 and 75 kev peaks. The modulation curves for 0° and 90° polarization angles are shown in the right column of Fig. 7, along with the modulation patterns obtained from simulation for different polarization fractions. Modulation factors are found to be low ($\sim 0.20$) indicating the beam is $\sim 40$ % polarized.

To further test the polarimetric performance of CXPOL at lower energies, we used an X-ray gun (Amptek Mini-X X-Ray Tube with Gold target) to obtain a continuum polarized beam. The X-ray gun emits in 10−50 kev range. The lower energy photons ($\leq 20$ kev) are blocked using a thin aluminium filter. We employed a similar method to polarize the continuum emission from the X-ray gun. An aluminium scatterer inside a lead cylinder was kept at the end of the gun tube (see Fig. 8). A small opening of 2 mm diameter ensures that the photons scattered at 90° can only reach the plastic scatterer. Spectrum of the polarized beam as taken from CdTe detector is shown in Fig. 8. We see the beam is continuum in 20−50 kev range. Because of the constrained geometry, we expect a higher degree of polarization in this case. Modulation curves at 0°, 90° and 45° polarization angles are shown in Fig. 9. In simulation, we employed a broad Gaussian beam centered around 35 kev as the source of polarized radiation. Azimuthal angle distributions for all the polarization planes are found to be highly modulated consistent with $\geq 90$ % polarization. All the experiments were repeated few times to have confidence in the obtained results. The coincidence time window in the experiments was set to $\sim 6$ $\mu$s, it can be altered to a smaller window of $\sim 3$ $\mu$s in order to further improve the signal to noise ratio.

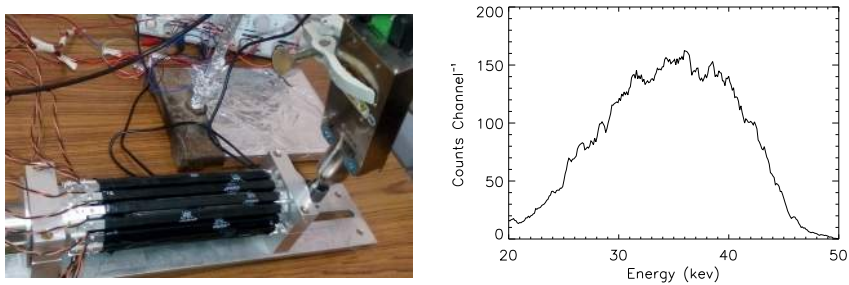

**Fig. 8** Left: polarization experiment with CXPOL using continuum radiation from X-ray gun. Right : Spectrum of the polarized 90° scattered radiation of the gun as taken from CdTe detector.

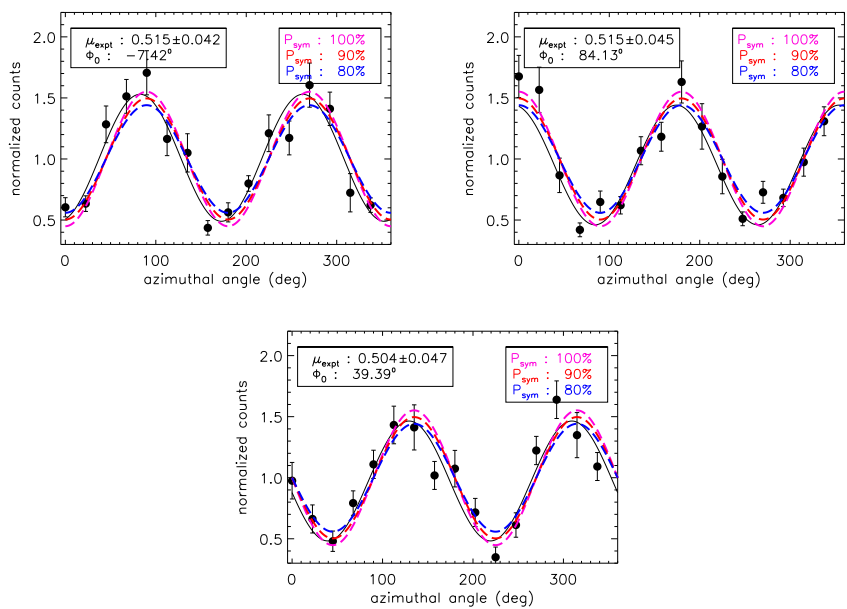

**Fig. 9** Azimuthal angle distribution for partially polarized 20−50 kev continuum radiation for 0° (top left), 90° (top right), and 45° (bottom) polarization angles. The black solid line is the fit to the experimental data. The pink, red and blue dashed lines represent the modulation curves obtained from simulation for this setup for 100, 90 and 80 % polarized beams respectively.

## 5 Discussions and future plans

The main highlights and importance of the experimental results are discussed below.

- SiPMs have been successfully implemented to read out CsI(Tl) absorbers for the Compton polarimeter, which helps in designing a compact and optimized polarimeter geometry.
- Proper choice of scintillator is extremely important in case of SiPM readout to have better polarimetric sensitivity. We showed that for CsI(Tl) scintillators viewed by a single SiPM, detection probability degrades significantly at a distance of 5−6 cm from SiPM at energies ∼ 50 kev. This is mainly due to longer scintillation time constant of CsI(Tl) scintillator as a result of which photons reaching SiPM simultaneously are less in number and therefore pulse amplitude is low. However, with $CeBr_3$, $LaBr_3(Ce)$, NaI(Tl), or other new generation scintillators, which has comparatively smaller decay time constant and at the same time similar or even higher light output, much better performance is expected.
- Choice of appropriate SiPM is also crucial to achieve better sensitivity at low energies. For lower threshold of the scintillators, SiPMs with less background (with similar active area) will be useful. Here we used SiPMs procured from KETEK, which has typical background level of $\leq 500$ kHz/mm$^2$. The new generation SiPMs are supposed to have comparatively much lower background. Recovery time of the micro-pixels after avalanche is also much smaller. These are the key factors to obtain a lower energy threshold of 20 kev or less for the Compton polarimeter. We plan to investigate the sensitivity of $CeBr_3$ and NaI(Tl) scintillators coupled to these new generation SiPMs in near future for the next version of the polarimeter. Coupling between the scintillator and SiPM is also a key factor for better performance of the scintillators. We plan for an optimized enclosure for the SiPM and scintillator which may lead to a better coupling between them and hence an improved threshold.
- With proper choice of scintillators and optimized scintillator-SiPM coupling, new generation SiPMs can be successfully used for lower energy detections ∼ 20 kev. This is encouraging as one can now think of new polarimetric configurations with SiPM in order to have simultaneous polarimetric and spectroscopic information.

  Use of two SiPMs at two ends of a scintillator will not only optimize the system for better energy threshold but will also give position of interaction from ratio of pulse heights. In that case, use of a Si detector in place of the central plastic scatterer will make the instrument sensitive for Compton spectroscopy in 20−80 kev apart from the high resolution spectroscopy up to ∼ 40 kev from Si alone. However, the polarimetric performance of the instrument will be compromised because of lower scattering efficiency of Si. On the other hand, with the use of a 5 mm thick plastic (viewed by SiPM array) below the SDD, it is possible to greatly improve the polarimetric performance of the system [49].

Another possible configuration is to use a central scatterer made of segmented plastics with each segment viewed by a SiPM and an array of scintillators with two sided SiPM readout. Such a configuration will optimize the instrument for polarimetry and Compton spectroscopy in $20-80$ kev. The overall energy resolution of the system would be limited by resolution of the absorbers and uncertainties in constraining the positions of interactions. However, compared to the Si scatterer configuration, the disadvantage of this configuration is the infeasibility of photoelectric spectroscopy at lower energies.

## 6 Summary

We are developing a proof of concept laboratory model of a hard X-ray focal plane Compton polarimeter, as a foundation for the future proposal of a dedicated hard X-ray polarimetry mission. The main objective here is to demonstrate a mature readiness level of a robust polarimeter configuration and to obtain firm estimates of the resources requirements (in terms of size, weight, power etc) for the future space experiment. In this paper, we demonstrated the characteristics of CsI(Tl)-SiPM system used to record the azimuthal angle distribution. We could successfully assemble full polarimeter configuration and test it with both unpolarized as well as partially polarized X-rays. The results presented in this paper are expected to be very useful for designing of future Compton polarimetry experiments with SiPM scintillator read out systems. While this is only a first version of the proposed polarimeter configuration, we could identify few issues with the initial design, which we plan to rectify in the subsequent versions.

**Acknowledgements** Research at Physical Research Laboratory, Ahmedabad is supported by the Department of Space, Government of India.

## References

1. Angel, J.R., Novick, R., vanden Bout, P., Wolff, R.: Search for X-Ray Polarization in Sco X-1. Physical Review Letters **22**, 861–865 (1969). DOI 10.1103/PhysRevLett.22.861
2. Beilicke, M., Kislat, F., Zajczyk, A., Guo, Q., Endsley, R., Stork, M., Cowsik, R., Dowkontt, P., Barthelmy, S., Hams, T., Okajima, T., Sasaki, M., Zeiger, B., de Geronimo, G., Baring, M.G., Krawczynski, H.: Design and Performance of the X-ray Polarimeter X-Calibur. Journal of Astronomical Instrumentation **3**, 1440008 (2014). DOI 10.1142/S225117171440008X
3. Bellazzini, R., Angelini, F., Baldini, L., Bitti, F., Brez, A., Cavalca, F., Del Prete, M., Kuss, M., Latronico, L., Omodei, N., Pinchera, M., Massai, M.M., Minuti, M., Razzano, M., Sgro, C., Spandre, G., Tenze, A., Costa, E., Soffitta, P.: Gas pixel detectors for X-ray polarimetry applications. Nuclear Instruments and Methods in Physics Research A **560**, 425–434 (2006). DOI 10.1016/j.nima.2006.01.046
4. Bellazzini, R., et al.: A polarimeter for IXO. In: R. Bellazzini, E. Costa, G. Matt, G. Tagliaferri (eds.) X-ray Polarimetry: A New Window in Astrophysics by Ronaldo Bellazzini, Enrico Costa, Giorgio Matt and Gianpiero Tagliaferri. Cambridge University Press, 2010. ISBN: 9780521191845, p. 269, p. 269 (2010)

5. Bellazzini, R., Spandre, G., Minuti, M., Baldini, L., Brez, A., Latronico, L., Omodei, N., Razzano, M., Massai, M.M., Pesce-Rollins, M., Sgró, C., Costa, E., Soffitta, P., Sipila, H., Lempinen, E.: A sealed Gas Pixel Detector for X-ray astronomy. Nuclear Instruments and Methods in Physics Research A **579**, 853–858 (2007). DOI 10.1016/j.nima.2007.05.304
6. Black, J.K., Baker, R.G., Deines-Jones, P., Hill, J.E., Jahoda, K.: X-ray polarimetry with a micropattern TPC. Nuclear Instruments and Methods in Physics Research A **581**, 755–760 (2007). DOI 10.1016/j.nima.2007.08.144
7. Bloser, P.F., Legere, J., Bancroft, C., McConnell, M.L., Ryan, J.M., Schwadron, N.: Scintillator gamma-ray detectors with silicon photomultiplier readouts for high-energy astronomy. In: Society of Photo-Optical Instrumentation Engineers (SPIE) Conference Series, *Society of Photo-Optical Instrumentation Engineers (SPIE) Conference Series*, vol. 8859, p. 0 (2013). DOI 10.1117/12.2024411
8. Bloser, P.F., Legere, J.S., McConnell, M.L., Macri, J.R., Bancroft, C.M., Connor, T.P., Ryan, J.M.: Calibration of the Gamma-RAy Polarimeter Experiment (GRAPE) at a polarized hard X-ray beam. Nuclear Instruments and Methods in Physics Research A **600**, 424–433 (2009). DOI 10.1016/j.nima.2008.11.118
9. Buzhan, P., Dolgoshein, B., Filatov, L., Ilyin, A., Kantzerov, V., Kaplin, V., Karakash, A., Kayumov, F., Klemin, S., Popova, E., Smirnov, S.: Silicon photomultiplier and its possible applications. Nuclear Instruments and Methods in Physics Research A **504**, 48–52 (2003). DOI 10.1016/S0168-9002(03)00749-6
10. Buzhan, P., Dolgoshein, B., Ilyin, A., Kantserov, V., Kaplin, V., Karakash, A., Pleshko, A., Popova, E., Smirnov, S., Volkov, Y., Filatov, L., Klemin, S., Kayumov, F.: The Advanced Study of Silicon Photomultiplier. In: M. Barone, E. Borchi, J. Huston, C. Leroy, P.G. Rancoita, P. Riboni, R. Ruchti (eds.) Advanced Technology - Particle Physics, pp. 717–728 (2002). DOI 10.1142/97898127764640101
11. Chattopadhyay, T., Vadawale, S.V., Pendharkar, J.: Compton polarimeter as a focal plane detector for hard X-ray telescope: sensitivity estimation with Geant4 simulations. Experimental Astronomy **35**, 391–412 (2013). DOI 10.1007/s10686-012-9312-3
12. Chattopadhyay, T., Vadawale, S.V., Rao, A.R., Sreekumar, S., Bhattacharya, D.: Prospects of hard X-ray polarimetry with Astrosat-CZTI. Experimental Astronomy **37**, 555–577 (2014). DOI 10.1007/s10686-014-9386-1
13. Chattopadhyay, T., Vadawale, S.V., Shanmugam, M., Goyal, S.K.: Measurement of Low Energy Detection Efficiency of a Plastic Scintillator: Implications on the Lower Energy Limit and Sensitivity of a Hard X-Ray Focal Plane Compton Polarimeter. Astrophysical Journal Supplement **212**, 12 (2014). DOI 10.1088/0067-0049/212/1/12
14. Coburn, W., Boggs, S.E.: Polarization of the prompt $\gamma$-ray emission from the $\gamma$-ray burst of 6 December 2002. Nature **423**, 415–417 (2003). DOI 10.1038/nature01612
15. Costa, E., Bellazzini, R., Tagliaferri, G., Matt, G., Argan, A., Attinà, P., Baldini, L., Basso, S., Brez, A., Citterio, O., di Cosimo, S., Cotroneo, V., Fabiani, S., Feroci, M., Ferri, A., Latronico, L., Lazzarotto, F., Minuti, M., Morelli, E., Muleri, F., Nicolini, L., Pareschi, G., di Persio, G., Pinchera, M., Razzano, M., Reboa, L., Rubini, A., Salonico, A.M., Sgro', C., Soffitta, P., Spandre, G., Spiga, D., Trois, A.: POLARIX: a pathfinder mission of X-ray polarimetry. Experimental Astronomy **28**, 137–183 (2010). DOI 10.1007/s10686-010-9194-1
16. Costa, E., Soffitta, P., Bellazzini, R., Brez, A., Lumb, N., Spandre, G.: An efficient photoelectric X-ray polarimeter for the study of black holes and neutron stars. Nature **411**, 662–665 (2001)
17. Dean, A.J., Clark, D.J., Stephen, J.B., McBride, V.A., Bassani, L., Bazzano, A., Bird, A.J., Hill, A.B., Shaw, S.E., Ubertini, P.: Polarized Gamma-Ray Emission from the Crab. Science **321**, 1183– (2008). DOI 10.1126/science.1149056
18. Elsner, R.F., Ramsey, B.D., O'dell, S.L., Sulkanen, M., Tennant, A.F., Weisskopf, M.C., Gunji, S., Minamitani, T., Austin, R.A., Kolodziejczak, J., Swartz, D., Garmire, G., Meszaros, P., Pavlov, G.G.: The X-ray Polarimeter Experiment (XPE). In: American Astronomical Society Meeting Abstracts #190, *Bulletin of the American Astronomical Society*, vol. 29, p. 790 (1997)
19. Forot, M., Laurent, P., Grenier, I.A., Gouiffès, C., Lebrun, F.: Polarization of the Crab Pulsar and Nebula as Observed by the INTEGRAL/IBIS Telescope. Astrophysical Journal Letter **688**, L29–L32 (2008). DOI 10.1086/593974

20. Götz, D., Covino, S., Fernández-Soto, A., Laurent, P., Bošnjak, Ž.: The polarized gamma-ray burst GRB 061122. MNRAS **431**, 3550–3556 (2013). DOI 10.1093/mnras/stt439
21. Götz, D., Laurent, P., Lebrun, F., Daigne, F., Bošnjak, Ž.: Variable Polarization Measured in the Prompt Emission of GRB 041219A Using IBIS on Board INTEGRAL. Astrophysical Journal Letter **695**, L208–L212 (2009). DOI 10.1088/0004-637X/695/2/L208
22. Gowen, R.A., Cooke, B.A., Griffiths, R.E., Ricketts, M.J.: An upper limit to the linear X-ray polarization of SCO X-1. MNRAS **179**, 303–310 (1977)
23. Griffiths, R.E., Ricketts, M.J., Cooke, B.A.: Observations of the X-ray nova A0620-00 with the Ariel V crystal spectrometer/polarimeter. MNRAS **177**, 429–440 (1976)
24. Hayashida, K., Yonetoku, D., Gunji, S., Tamagawa, T., Mihara, T., Mizuno, T., Takahashi, H., Dotani, T., Kubo, H., Yatsu, Y., Tokanai, F., Nakamori, T., Shibata, S., Hayato, A., Furuzawa, A., Kishimoto, Y., Kitamoto, S., Toma, K., Sadamoto, M., Yoshinaga, K., Kim, J., Ide, S., Kamitsukasa, F., Anabuki, N., Tsunemi, H., Katagiri, J., Sugimoto, J.: X-ray gamma-ray polarimetry small satellite PolariS. In: Society of Photo-Optical Instrumentation Engineers (SPIE) Conference Series, *Society of Photo-Optical Instrumentation Engineers (SPIE) Conference Series*, vol. 9144, p. 0 (2014). DOI 10.1117/12.2056685
25. Hughes, J.P., Long, K.S., Novick, R.: A search for X-ray polarization in cosmic X-ray sources. Astrophysical Journal **280**, 255–258 (1984). DOI 10.1086/161992
26. Jahoda, K.: The Gravity and Extreme Magnetism Small Explorer. In: Society of Photo-Optical Instrumentation Engineers (SPIE) Conference Series, *Society of Photo-Optical Instrumentation Engineers (SPIE) Conference Series*, vol. 7732 (2010). DOI 10.1117/12.857439
27. Jourdain, E., Roques, J.P., Chauvin, M., Clark, D.J.: Separation of Two Contributions to the High Energy Emission of Cygnus X-1: Polarization Measurements with INTEGRAL SPI. Astrophysical Journal **761**, 27 (2012). DOI 10.1088/0004-637X/761/1/27
28. Kaaret, P., Novick, R., Martin, C., Hamilton, T., Sunyaev, R., Lapshov, I., Silver, E., Weisskopf, M., Elsner, R., Chanan, G., Manzo, G., Costa, E., Fraser, G., Perola, G.C.: SXRP. A focal plane stellar X-ray polarimeter for the SPECTRUM-X-Gamma mission. In: R.B. Hoover (ed.) X-Ray/EUV Optics for Astronomy and Microscopy, *Society of Photo-Optical Instrumentation Engineers (SPIE) Conference Series*, vol. 1160, pp. 587–597 (1989)
29. Kamae, T., Andersson, V., Arimoto, M., Axelsson, M., Marini Bettolo, C., Björnsson, C.I., Bogaert, G., Carlson, P., Craig, W., Ekeberg, T., Engdegård, O., Fukazawa, Y., Gunji, S., Hjalmarsdotter, L., Iwan, B., Kanai, Y., Kataoka, J., Kawai, N., Kazejev, J., Kiss, M., Klamra, W., Larsson, S., Madejski, G., Mizuno, T., Ng, J., Pearce, M., Ryde, F., Suhonen, M., Tajima, H., Takahashi, H., Takahashi, T., Tanaka, T., Thurston, T., Ueno, M., Varner, G., Yamamoto, K., Yamashita, Y., Ylinen, T., Yoshida, H.: PoGOLite A high sensitivity balloon-borne soft gamma-ray polarimeter. Astroparticle Physics **30**, 72–84 (2008). DOI 10.1016/j.astropartphys.2008.07.004
30. Katsuta, J., Mizuno, T., Ogasaka, Y., Yoshida, H., Takahashi, H., Kano, Y., Iwahara, T., Sasaki, N., Kamae, T., Kokubun, M., Takahashi, T., Hayashida, K., Uesugi, K.: Evaluation of polarization characteristics of multilayer mirror for hard X-ray observation of astrophysical objects. Nuclear Instruments and Methods in Physics Research A **603**, 393–400 (2009). DOI 10.1016/j.nima.2009.02.039
31. Kishimoto, Y., Gunji, S., Ishigaki, Y., Kanno, M., Murayama, H., Ito, C., Tokanai, F., Suzuki, K., Sakurai, H., Mihara, T., Kohama, M., Suzuki, M., Hayato, A., Hayashida, K., Anabuki, N., Morimoto, M., Tsunemi, H., Saito, Y., Yamagami, T., Kishimoto, S.: Basic Performance of PHENEX: A Polarimeter for High ENErgy X rays. IEEE Transactions on Nuclear Science **54**, 561–566 (2007). DOI 10.1109/TNS.2007.897827
32. Kunieda, H., Awaki, H., Furuzawa, A., Haba, Y., Iizuka, R., Ishibashi, K., Ishida, M., Itoh, M., Kosaka, T., Maeda, Y., Matsumoto, H., Miyazawa, T., Mori, H., Namba, Y., Ogasaka, Y., Ogi, K., Okajima, T., Suzuki, Y., Tamura, K., Tawara, Y., Uesugi, K., Yamashita, K., Yamauchi, S.: Hard x-ray telescope to be onboard ASTRO-H. In: Society of Photo-Optical Instrumentation Engineers (SPIE) Conference Series, *Society of Photo-Optical Instrumentation Engineers (SPIE) Conference Series*, vol. 7732 (2010). DOI 10.1117/12.856892

33. Laurent, P., Rodriguez, J., Wilms, J., Cadolle Bel, M., Pottschmidt, K., Grinberg, V.: Polarized Gamma-Ray Emission from the Galactic Black Hole Cygnus X-1. Science **332**, 438– (2011). DOI 10.1126/science.1200848
34. Marshall, H.L., Murray, S.S., Chappell, J.H., Schnopper, H.W., Silver, E.H., Weisskopf, M.C.: Realistic, inexpensive, soft x-ray polarimeter and the potential scientific return. In: S. Fineschi (ed.) Polarimetry in Astronomy, *Society of Photo-Optical Instrumentation Engineers (SPIE) Conference Series*, vol. 4843, pp. 360–371 (2003). DOI 10.1117/12.459486
35. McConnell, M.L., Ryan, J.M., Smith, D.M., Lin, R.P., Emslie, A.G.: RHESSI as a Hard X-Ray Polarimeter. Solar Physics **210**, 125–142 (2002). DOI 10.1023/A:1022413708738
36. McGlynn, S., Clark, D.J., Dean, A.J., Hanlon, L., McBreen, S., Willis, D.R., McBreen, B., Bird, A.J., Foley, S.: Polarisation studies of the prompt gamma-ray emission from GRB 041219a using the spectrometer aboard INTEGRAL. Astronomy & Astrophysics **466**, 895–904 (2007). DOI 10.1051/0004-6361:20066179
37. McGlynn, S., Foley, S., McBreen, B., Hanlon, L., McBreen, S., Clark, D.J., Dean, A.J., Martin-Carrillo, A., O'Connor, R.: High energy emission and polarisation limits for the INTEGRAL burst GRB 061122. Astronomy & Astrophysics **499**, 465–472 (2009). DOI 10.1051/0004-6361/200810920
38. Moran, P., Shearer, A., Gouiffes, C., Laurent, P.: INTEGRAL/IBIS and optical observations of the Crab nebula/pulsar polarisation. ArXiv e-prints (2013)
39. Novick, R., Weisskopf, M.C., Berthelsdorf, R., Linke, R., Wolff, R.S.: Detection of X-Ray Polarization of the Crab Nebula. Astrophysical Journal Letter **174**, L1 (1972). DOI 10.1086/180938
40. Orsi, S., Polar Collaboration: POLAR: A Space-borne X-Ray Polarimeter for Transient Sources. Astrophysics and Space Sciences Transactions **7**, 43–47 (2011). DOI 10.5194/astra-7-43-2011
41. Otte, N.: The Silicon Photomultiplier-a new device for high energy physics, astroparticle physics, industrial and medical applications. In: IX International Symposium on Detectors for Particle, Astroparticle and Synchrotron Radiation Experiments, SNIC Symposium, Stanford, California, pp. 1–9 (2006)
42. Rishin, P.V., et al.: Development of a Thomson X-ray polarimeter. In: R. Bellazzini, E. Costa, G. Matt, G. Tagliaferri (eds.) X-ray Polarimetry: A New Window in Astrophysics by Ronaldo Bellazzini, Enrico Costa, Giorgio Matt and Gianpiero Tagliaferri. Cambridge University Press, 2010. ISBN: 9780521191845, p. 83, p. 83 (2010)
43. Sanaei, B., Baei, M.T., Sayyed-Alangi, S.Z.: Characterization of a New Silicon Photomultiplier in Comparison with a Conventional Photomultiplier Tube. Journal of Modern Physics **6**, 425–433 (2015). DOI 10.4236/jmp.2015.64046
44. Silver, E.H., Weisskopf, M.C., Kestenbaum, H.L., Long, K.S., Novick, R., Wolff, R.S.: The first search for X-ray polarization in the Centaurus X-3 and Hercules X-1 pulsars. Astrophysical Journal **232**, 248–254 (1979). DOI 10.1086/157283
45. Soffitta, P., Barcons, X., Bellazzini, R., Braga, J., Costa, E., Fraser, G.W., Gburek, S., Huovelin, J., Matt, G., Pearce, M., Poutanen, J., Reglero, V., Santangelo, A., Sunyaev, R.A., Tagliaferri, G., Weisskopf, M., Aloisio, R., Amato, E., Attiná, P., Axelsson, M., Baldini, L., Basso, S., Bianchi, S., Blasi, P., Bregeon, J., Brez, A., Bucciantini, N., Burderi, L., Burwitz, V., Casella, P., Churazov, E., Civitani, M., Covino, S., Curado da Silva, R.M., Cusumano, G., Dadina, M., D'Amico, F., De Rosa, A., Di Cosimo, S., Di Persio, G., Di Salvo, T., Dovciak, M., Elsner, R., Eyles, C.J., Fabian, A.C., Fabiani, S., Feng, H., Giarrusso, S., Goosmann, R.W., Grandi, P., Grosso, N., Israel, G., Jackson, M., Kaaret, P., Karas, V., Kuss, M., Lai, D., Rosa, G.L., Larsson, J., Larsson, S., Latronico, L., Maggio, A., Maia, J., Marin, F., Massai, M.M., Mineo, T., Minuti, M., Moretti, E., Muleri, F., O'Dell, S.L., Pareschi, G., Peres, G., Pesce, M., Petrucci, P.O., Pinchera, M., Porquet, D., Ramsey, B., Rea, N., Reale, F., Rodrigo, J.M., Różańska, A., Rubini, A., Rudawy, P., Ryde, F., Salvati, M., de Santiago, V.A., Sazonov, S., Sgró, C., Silver, E., Spandre, G., Spiga, D., Stella, L., Tamagawa, T., Tamborra, F., Tavecchio, F., Teixeira Dias, T., van Adelsberg, M., Wu, K., Zane, S.: XIPE: the X-ray imaging polarimetry explorer. Experimental Astronomy **36**, 523–567 (2013). DOI 10.1007/s10686-013-9344-3
46. Soffitta, P., Costa, E., Kaaret, P., Dwyer, J., Ford, E., Tomsick, J., Novick, R., Nenonen, S.: Proportional counters for the stellar X-ray polarimeter with a wedge and strip cath-ode pattern readout system. Nuclear Instruments and Methods in Physics Research A **414**, 218–232 (1998). DOI 10.1016/S0168-9002(98)00572-5

47. Tagliaferri, G., Argan, A., Bellazzini, R., Bookbinder, J., Catalano, O., Cavazzuti, E., Costa, E., Cusumano, G., Fiore, F., Fiorini, C., Giommi, P., Malaguti, G., Matt, G., Mereghetti, S., Micela, G., Murray, S., Negri, B., Pareschi, G., Perola, G., Romaine, S., Villa, G.: NHXM: a New Hard X-ray imaging and polarimetric Mission. In: Society of Photo-Optical Instrumentation Engineers (SPIE) Conference Series, *Society of Photo-Optical Instrumentation Engineers (SPIE) Conference Series*, vol. 7732, p. 17 (2010). DOI 10.1117/12.857032
48. Tsunemi, H., Hayashida, K., Tamura, K., Nomoto, S., Wada, M., Hirano, A., Miyata, E.: Detection of X-ray polarization with a charge coupled device. Nuclear Instruments and Methods in Physics Research A **321**, 629–631 (1992). DOI 10.1016/0168-9002(92)90075-F
49. Vadawale, S.V., Chattopadhyay, T., Pendharkar, J.: A conceptual design of hard X-ray focal plane detector for simultaneous x-ray polarimetric, spectroscopic, and timing measurements. In: Society of Photo-Optical Instrumentation Engineers (SPIE) Conference Series, *Society of Photo-Optical Instrumentation Engineers (SPIE) Conference Series*, vol. 8443 (2012). DOI 10.1117/12.935295
50. Vadawale, S.V., Chattopadhyay, T., Rao, A.R., Bhattacharya, D., Bhalerao, V.B., Vagshette, N., Pawar, P., Sreekumar, S.: Hard X-ray polarimetry with Astrosat-CZTI. Astronomy & Astrophysics **578** (2015). DOI 10.1051/0004-6361/201525686
51. Vadawale, S.V., Paul, B., Pendharkar, J., Naik, S.: Comparative study of different scattering geometries for the proposed Indian X-ray polarization measurement experiment using Geant4. Nuclear Instruments and Methods in Physics Research A **618**, 182–189 (2010). DOI 10.1016/j.nima.2010.02.116
52. Weisskopf, M.C., Cohen, G.G., Kestenbaum, H.L., Long, K.S., Novick, R., Wolff, R.S.: Measurement of the X-ray polarization of the Crab Nebula. Astrophysical Journal Letter **208**, L125–L128 (1976). DOI 10.1086/182247
53. Weisskopf, M.C., Silver, E.H., Kestenbaum, H.L., Long, K.S., Novick, R.: A precision measurement of the X-ray polarization of the Crab Nebula without pulsar contamination. Astrophysical Journal Letter **220**, L117–L121 (1978). DOI 10.1086/182648
54. Wolff, R.S., Angel, J.R.P., Novick, R., vanden Bout, P.: Search for Polarization in the X-Ray Emission of the Crab Nebula. Astrophysical Journal Letter **160**, L21 (1970). DOI 10.1086/180515
55. Yatsu, Y., Ito, K., Kurita, S., Arimoto, M., Kawai, N., Matsushita, M., Kawajiri, S., Kitamura, S., Matunaga, S., Kimura, S., Kataoka, J., Nakamori, T., Kubo, S.: Preflight performance of a micro-satellite TSUBAME for X-ray polarimetry of gamma-ray bursts. In: Society of Photo-Optical Instrumentation Engineers (SPIE) Conference Series, *Society of Photo-Optical Instrumentation Engineers (SPIE) Conference Series*, vol. 9144, p. 0 (2014). DOI 10.1117/12.2056275
56. Yonetoku, D., Murakami, T., Masui, H., Kodaira, H., Aoyama, Y., Gunji, S., Tokanai, F., Mihara, T.: Development of polarimeter for gamma-ray bursts onboard the solar-powered sail mission. In: Society of Photo-Optical Instrumentation Engineers (SPIE) Conference Series, *Society of Photo-Optical Instrumentation Engineers (SPIE) Conference Series*, vol. 6266 (2006). DOI 10.1117/12.670134